\documentclass[prl,aps,twocolumn,epsfig,nofootinbib]{revtex4}
 \usepackage{CJK}
\usepackage{amsmath}
\usepackage{amssymb}
\usepackage[dvips]{graphicx}
\usepackage{dcolumn}
\usepackage{bm}
\usepackage[colorlinks=false,dvipdfm]{hyperref}
\usepackage{setspace}
\usepackage{amsfonts}
\usepackage{rotating}
\usepackage{makeidx}
\usepackage{amsthm}

\begin{document}

\title{Dynamical Algebras in the 1+1 Dirac Oscillator and the Jaynes--Cummings Model\footnote{ Chin. Phys. Lett.  37(5)  050301 (2020).}}
\author{Wen-Ya Song}
\affiliation{Department of Physics, School of Science, Tianjin University, Tianjin 300072, China}
\author{Fu-Lin Zhang\footnote{Corresponding author:  flzhang@tju.edu.cn}}
\affiliation{Department of Physics, School of Science, Tianjin University, Tianjin 300072, China}

\date{\today}

\begin{abstract}
 We study the algebraic structure  of the one-dimensional Dirac oscillator by extending the concept of spin symmetry to a noncommutative case.
 An $SO(4)$ algebra is found connecting the eigenstates of the Dirac oscillator, in which the two elements of Cartan subalgebra are conserved quantities.
 Similar results are obtained in the  Jaynes--Cummings model.                                                                                                                                                                                                            \end{abstract}

\maketitle

  The Dirac equation is a cornerstone of modern physics \cite{DiracQM}.
It is consistent with both the principles of quantum mechanics and the theory of special relativity,
and it plays an essential role in the field of high energy physics \cite{ZeeQFT}.
The Dirac oscillator (DO), which is obtained by the substitution ${\bm p}\rightarrow {\bm p}- {i} \beta m \omega {\bm x}$ into the free Dirac
equation \cite{moshinsky1989,RevDO2010}, has become the paradigm for the construction of covariant quantum models with some well determined
nonrelativistic limits \cite{PhysRevLett.111.170405}.
It is used in various branches of physics, such as nuclear physics \cite{PhysRevC.85.054617} and subnuclear physics \cite{PhysRevA.84.052102},
and it has recently been simulated in quantum optics \cite{PhysRevA.76.041801,PhysRevLett.99.123602,PhysRevLett.98.253005,AIPCP2010JC} and classical microwave
setups \cite{Sadurn__2010,PhysRevLett.111.170405}.
What is particularly noteworthy is that the simulation in quantum optics is based on
the equivalence between the DO and the Jaynes--Cummings (JC) model,
which is the simplest soluble model in light-matter interaction within the frame of completely quantized theory \cite{JC1963}.

The concepts of dynamical symmetry  and algebraic structure are essential and prevalent in both classical and quantum mechanics
\cite{GreinerSymm,OperMeth}.
In two simple examples, the Hydrogen atom and harmonic oscillator in nonrelativistic quantum mechanics \cite{GreinerSymm,OperMeth}, the energy levels of a quantum system may be derived based on its algebraic structure.
Algebraic properties of the DO has been studied  from different perspectives  \cite{lange1991,benitez1990} shortly after the original work of Moshinsky
\cite{moshinsky1989}.
Recently, Zhou {\it et al}. \cite{CPLZhouJ}  constructed shift operators
of the JC model\footnote{The words 'of the JC model' is missed out in the published version.}
using the matrix-diagonalizing technique \cite{ge2000unified}.

In this work, we re-examine the algebraic structures of  the one-dimensional (1D) DO and the JC model.
The motivation for this research is that the eigenstates of the JC model (and similarly, the DO) are determined by two good quantum numbers,
but only one pair of raising and lowering operators were obtained in Ref. \cite{CPLZhouJ}.
Therefore, our aim is to derive the two conserved quantities accompanied by their shift operators, which describe the full algebraic structure of the two models.

Our work begins with the 1D DO,
which can be obtained by linearizing the quadratic form $E^2=m^2+p^2+m^2\omega^2 x^2- \beta m \omega$ because original approach to derive the Dirac
Hamiltonian \cite{DiracQM},  where $\omega $  and $\beta$ are the frequency and one of the Dirac matrices.
We choose $\hbar=c=1$ in this study.
This implies that the 1D DO and the two-dimensional (2D) free Dirac system may share similar features, whereas difference comes from the commutator $[x,p]=i $.

The Dirac Hamiltonian, with scalar and vector potentials of equal magnitude, certainly including the free Dirac equation,  has the spin or pseudospin symmetry
corresponding to the same or opposite sign \cite{ginocchio2005relativistic}.
These systems are shown to have the same dynamical symmetries with their nonrelativistic  counterparts
\cite{ginocchio2004relativistic,ginocchio2005u,zhang2008dynamical,zhang2009dynamical,PhysRevA.80.054102}.
The key step to reveal the symmetries  is to derive the conserved  quantities with the aid of a unitary operator
\cite{ginocchio2004relativistic,ginocchio2005u,zhang2008dynamical}.
Zhang \textit{el al.} proposed a 2D version of the unitary operator \cite{zhang2009dynamical},
 and showed that it can be used to deal with  noncentral systems \cite{PhysRevA.80.054102}.

In this work,  we  extend the unitary for the 2D Dirac equation to a noncommutative case.
Namely, one of the momentums in the unitary is replaced by a coordinate operator.
Using this unitary, one can derive a conserved number operator and a spin of the 1D DO,i.e.,
the shift operators  which can be constructed simultaneously.
With these conserved quantities and shift operators, an $SO(4)$ algebraic structure of the system is revealed,   which is said to be a dynamical symmetry in the broad
sense \cite{chen1998application}.
Similar results are obtained in the  JC model by a similar procedure.

Let us begin with the conserved angular momentum of the 2D free Dirac equation.
The Hamiltonian reads
\begin{eqnarray}\label{H2D}
H={\bm\alpha}\cdot {\bm p}+\beta m,
\end{eqnarray}
where the Dirac matrices are conveniently defined in terms of the Pauli matrices ${\bm\alpha}=(\sigma_1,\sigma_2)$ and $\beta=\sigma_3$.
In the matrix form, it reads
\begin{align}\label{H2DM}
 {H}&=\begin{array}{cc}
 \left(
   \begin{array}{ccc}
    m & B^\dag \\
     B&-m\\
     \end{array}
 \right),
\end{array}
\end{align}
with $B=p_{1}-ip_{2}$ and $B^\dag =p_{1}+ip_{2}$.
The spin-orbit coupling leads to the fact that the usual orbital angular momentum $l=x_1 p_2 - x_2 p_1$ does not commute with the Hamiltonian (\ref{H2D}).
A deformed orbit momentum defined  in Ref. \cite{zhang2009dynamical}
\begin{eqnarray}
	L=U^\dag l U
\end{eqnarray}
can be easily proved to be conserved, when the unitary
\begin{eqnarray}\label{U2D}
U=\left(\begin{array}{cc}1&0\\0&\frac{B}{p}
\end{array}\right)
\end{eqnarray}
with   $p=\sqrt{p_1^2+p_2^2}=\sqrt{B^\dag B} = \sqrt{B B^\dag}$.
It equals
\begin{align}
L=UlU^\dag
=\begin{array}{cc}
 \left(
   \begin{array}{ccc}
     l & 0 \\
     0&l-1\\
     \end{array}
 \right).
\end{array}
\end{align}
and  commutes with the 2D Dirac Hamiltonian  with equal scalar and vector  radial potentials,
which is the key reason why the dynamical symmetries in these systems studied in Ref.\cite{zhang2009dynamical} are not broken
by the spin-orbit coupling.

Now, we turn to the conserved number operator of the 1D DO.
The Hamiltonian of the 1D version of the DO is given by
 \begin{equation}\label{HD}
\mathcal{H}= \alpha(p-{i}\beta m\omega x) +\beta m,
\end{equation}
where the Dirac matrices $\alpha =\sigma_2$ and $\beta=\sigma_3$.
In the matrix form, it reads
\begin{align}\label{HDOM}
\mathcal{H}&=\begin{array}{cc}
 \left(
   \begin{array}{ccc}
    m & \sqrt{2m\omega}a^\dag \\
     \sqrt{2m\omega}a&-m\\
     \end{array}
 \right),
\end{array}
\end{align}
where $a^\dag =\sqrt{\frac{m\omega}{2}}(x-\frac{i}{m\omega}p)$ and $a =\sqrt{\frac{m\omega}{2}}(x+\frac{i}{m\omega}p) $ are
the raising and lowering operators of  a nonrelativistic harmonic oscillator.
The 1D DO can be regarded as a coupled quantum system composed by the harmonic oscillator and a spin,
and the coupling vanishes  in the nonrelativistic limit \cite{cao2019analytic}.
This interaction  leads to the case that the number operator $N=a^\dag a$, which is a good quantum number of the harmonic oscillator, is no longer conserved.
 It is noteworthy that the matrix (\ref{HDOM}) can be obtained by replacing the first momentum with the coordinate corresponding to the second momentum.
This similarity between the Hamiltonians enlightens us
to construct a deformed number operator which commutes with the DO Hamiltonian.
Our solution is to extend the unitary  (\ref{U2D}) to the case of the DO.
An intuitive approach is to replace $B$ and $B^\dag$ in (\ref{U2D}) with the shift operators $a$ and $a^\dag$.
However, there are several possible alternatives because of the uncommutation between $a$ and $a^\dag$, or equivalently $x$ and $p$.
Among them, one can directly find the unitary
\begin{eqnarray}\label{UDO}
\mathcal{U}=\left(\begin{array}{cc}1&0\\0&a\frac{1}{\sqrt{N}}
\end{array}\right)
\end{eqnarray}
transforming the  number operator into a conserved one as
\begin{align}\label{Ndef}
    \mathcal{N}&=\mathcal{U}N\mathcal{U}^\dag=\begin{array}{cc}
 \left(
   \begin{array}{ccc}
     N& 0 \\
     0&N+1\\
     \end{array}
 \right).
\end{array}
\end{align}
It commutes with the Hamiltonian $\mathcal{H}$.
In addition, the spin operator is invariant  under the unitary; that is,
\begin{eqnarray}
s=\mathcal{U}s\mathcal{U}^\dag,
\end{eqnarray}
where  $s=\frac{\sigma_3}{2}$.

The spectrum of the 1D DO can be derived in the following two steps.
Under the inverse transformation of the unitary (\ref{UDO}), the Hamiltonian becomes
\begin{align}
 \mathcal{U}^\dag  \mathcal{H} \mathcal{U}=\begin{array}{cc}
 \left(
   \begin{array}{ccc}
     m& \sqrt{2m\omega N}  \\
     \sqrt{2m\omega N}  &-m\\
     \end{array}
 \right).
\end{array}
\end{align}
It can be diagonalized in the two-dimensional subspace with the same quantum number $N$; that is,
\begin{eqnarray}\label{HDOD}
e^{i\frac{\sigma_{2}}{2}\theta_N}  \mathcal{U}^\dag  \mathcal{H} \mathcal{U} e^{-i\frac{\sigma_{2}}{2}\theta_N}   =\sqrt{2m\omega N+m^{2}}\sigma_{3},
\end{eqnarray}
where $\theta_N$ is an operator function defined by $\tan\theta_N= \sqrt{ {2 \omega N }/{m}}$.
Here, we remark that the above form in (\ref{HDOD}) is different from the Foldy--Wouthuysen transformation \cite{moreno1989covariance}.

The eigenstates of the DO can be obtained directly as
\begin{eqnarray}
|\varphi_{n}^{\pm}\rangle = \mathcal{U} e^{-i\frac{\sigma_{2}}{2}\theta_N} |\pm\rangle\otimes | n\rangle,
\end{eqnarray}
where $ |\pm\rangle$ and $ | n\rangle$  are eigenstates of the spin and the harmonic oscillator, respectively,
satisfying $\sigma_3  |\pm\rangle= \pm|\pm\rangle$  and  $ N| n\rangle= n | n\rangle$.
Here, the quantum number $n=0,1,2 ,\ldots $ for the plus sign ($+$), but  $n=1,2 ,\ldots $ for the minus sign ($-$).
They satisfy
\begin{eqnarray}
\mathcal{H}|\varphi_{n}^{\pm}\rangle  =\mathcal{E}^{\pm}_n  |\varphi_{n}^{\pm}\rangle,
\end{eqnarray}
and the corresponding eigenenergies are
\begin{eqnarray}
\mathcal{E}^{\pm}_n=\pm \sqrt{2m\omega n+m^{2}}.
\end{eqnarray}
The eigenstates in the matrix form are given by
\begin{eqnarray}
|\varphi_{n}^{+}\rangle&\!\! =\!\!
 \left(\!
   \begin{array}{ccc}
  \cos\frac{\theta_n}{2}|n\rangle \\
  \sin\frac{\theta_n}{2} |n\!  -\!  1\rangle\\
     \end{array}
 \!\right)\! , \
|\varphi_{n}^{-}\rangle&\!\!  = \!\!
 \left(\!
   \begin{array}{ccc}
  -\sin\frac{\theta_n}{2}|n\rangle \\
 \cos\frac{\theta_n}{2} |n\!  -\!  1\rangle\\
     \end{array}
 \! \right)\!,
\end{eqnarray}
where $\theta_n$ is the eigenvalue of $\theta_N$ satisfying $\tan\theta_n= \sqrt{ {2 \omega n }/{m}}$.

Now we investigate the full algebraic structure of the 1D DO.
The diagonalized form of the Hamiltonian (\ref{HDOD}) indicates that
 there is a conserved quantity in addition to the deformed number operator,
 which is given by
  \begin{eqnarray}
\varSigma_3 =    \mathcal{U} e^{-i\frac{\sigma_{2}}{2}\theta_N} \sigma_3  e^{i\frac{\sigma_{2}}{2}\theta_N}  \mathcal{U}^\dag.
  \end{eqnarray}
The deformed number operator (\ref{Ndef}) equals
   \begin{eqnarray}
 \mathcal{N} =    \mathcal{U} e^{-i\frac{\sigma_{2}}{2}\theta_N} N  e^{i\frac{\sigma_{2}}{2}\theta_N}  \mathcal{U}^\dag,
   \end{eqnarray}
as $[ N,\theta_N]=0$.
The shift operators of the two conserved quantities can be obtained as
    \begin{eqnarray}
&b  \! = \!   \mathcal{U} e^{-i\frac{\sigma_{2}}{2}\theta_N} \!  a  e^{i\frac{\sigma_{2}}{2}\theta_N}  \mathcal{U}^\dag,~~
b^\dag  \! =  \!  \mathcal{U} e^{-i\frac{\sigma_{2}}{2}\theta_N} \! a^\dag\!   e^{i\frac{\sigma_{2}}{2}\theta_N}  \mathcal{U}^\dag, & \ \ \ \\
 &\varSigma_{\pm} =    \mathcal{U} e^{-i\frac{\sigma_{2}}{2}\theta_N} \sigma_{\pm}  e^{i\frac{\sigma_{2}}{2}\theta_N}  \mathcal{U}^\dag, &
    \end{eqnarray}
 where $\sigma_{\pm}= (\sigma_{1} \pm i\sigma_{2})/2$.
They satisfy the commutation relations of the harmonic oscillator and the Pauli matrices as
\begin{eqnarray}
 &[ \mathcal{N},b]=-b,~~[ \mathcal{N},b^\dag]=b^\dag,    ~~ [b, b^\dag] =1, & \\
 &[\varSigma_3,\varSigma_{\pm} ] = \pm 2\varSigma_{\pm},~~  [\varSigma_+,\varSigma_-] =  \varSigma_3. &
 \end{eqnarray}

The two conserved quantities and their shift operators constitute  two independent  Lie algebras.
The harmonic oscillator operators  construct an $SU(1,1)$ algebra  \cite{ge2000unified,CPLZhouJ} as
 \begin{eqnarray}
 K_{3}=\mathcal{N}+\frac{1}{2},~~~ K_{+}=b^{\dag}\xi(K_{3}),~~~ K_{-}=\xi(K_{3})b,
 \end{eqnarray}
with $\xi(K_{3})=\sqrt{\mathcal{N}+1}$,
and they satisfy the commutation relations
\begin{eqnarray}
	[K_{3},K_{\pm}]=\pm K_{\pm},~~ [K_{+},K_{-}]=- 2K_{3}.
 \end{eqnarray}
 The deformed  Pauli matrices construct an $SU(2)$ algebra naturally as
  \begin{eqnarray}
 S_{3}=\frac{\varSigma_3}{2},~~ S_{\pm} =\varSigma_{\pm},
  \end{eqnarray}
satisfying
\begin{eqnarray}
	[S_{3},S_{\pm}]=\pm S_{\pm},~~ [S_{+},S_{-}]=2S_{3}.
 \end{eqnarray}
 The non-hermitian generators in the two Lie algebras can be rewritten as the Hermitian ones as follows
 \begin{eqnarray}
 K_1=\frac{1}{2}(K_+ + K_-),&~~K_2=\frac{1}{2i}(K_+ - K_-),\\
S_1=\frac{1}{2}(S_+ + S_-),&~~ S_2=\frac{1}{2i}(S_+ - S_-).
  \end{eqnarray}
The two algebras are decoupled, as
\begin{eqnarray}
	[S_{i},K_{j}]=0,
 \end{eqnarray}
 with $i,j=1,2,3$.

 These results show that the 1D DO has an $SO(4)$ algebraic structure,
and its six generators are defined as
\begin{eqnarray}
I_i=\tilde{K}_i+S_i,~~  R_i=\tilde{K}_i-S_i,~~ \mathrm{with}~ i=1,2,3 ,
\end{eqnarray}
where $\tilde{K}_1=i K_1$, $\tilde{K}_2=i K_2$, $\tilde{K}_3= K_3$.
The commutation relations of the $SO(4)$ algebra are
\begin{eqnarray}
&&[I_i,I_j] = i \epsilon_{ijk} I_k, \nonumber\\
 &&[I_i,R_j] = i \epsilon_{ijk} R_k,\nonumber\\
 &&[R_i,R_j] = i \epsilon_{ijk} I_k,
\end{eqnarray}
 with $i,j,k=1,2,3$.
 The two conserved Hermitian operators $I_3$ and $R_3$,  satisfying  $[I_3, R_3]=0$, form the Catan subalgebra.
 The Hilbert space of the 1D DO provides a  representation of the $SO(4)$ Lie algebra, in which the $I_3$ and $R_3$ are diagonal.
 One can derive their matrix elements directly as
 \begin{eqnarray}
 && \!\! \!\! \!\! A_1 |\varphi^{\pm}_n\rangle \!= \!\frac{i}{2}[(n\!+\!1)|\varphi^{\pm}_{n+1}\rangle \!+\! n|\varphi^{\pm}_{n-1}\rangle]\! +\! (\!-\!1\!)^{\alpha} \frac{1}{2}|\varphi^{\mp}_n\rangle,  \nonumber\\
 && \!\! \!\! \!\! A_2 |\varphi^{\pm}_n\rangle \!= \! \frac{1}{2}[(n\!+\!1)|\varphi^{\pm}_{n+1}\rangle \!-\! n|\varphi^{\pm}_{n-1}\rangle]\! \pm\! (\!-\!1\!)^{\alpha} \frac{i}{2} |\varphi^{\mp}_n\rangle, \ \ \  \nonumber\\
 && \!\! \!\! \!\! A_3 |\varphi^{\pm}_n\rangle \!= \!  [n\!+\!1 \pm(\!-\!1\!)^{\alpha} \frac{1}{2}] |\varphi^{\pm}_n\rangle ,
 \end{eqnarray}
where $A=I$ with  $\alpha=0$, and $A=R$ with  $\alpha=1$.

We now turn to the JC model, the simplest completely quantized model in light-matter interaction,
in which a two-level atom (matter) is coupled with a quantized mode (light) of an optical cavity.
  This has many applications, not only in the field of  quantum optics \cite{vedral2005modern} but also in the field of solidstate quantum information circuits \cite{irish2007generalized}, both experimentally and theoretically.
 The Hamiltonian reads
 \begin{eqnarray}
\mathcal{H}_{JC}=\omega a^\dag a+\frac{\varOmega}{2}\sigma_{z}+J(a^\dag \sigma^{-}+a\sigma^{+}),
\end{eqnarray}
where $\varOmega$ is the level splitting of the two-level system, $a$ ($a^\dag$) is the destruction (creation) operator of a single bosonic mode with frequency $\omega$, $J$ is the coupled coefficient.
In the matrix form it is given by
\begin{eqnarray}
\mathcal{H}_{JC}=\left(\begin{array}{cc}\omega a^\dag a+\frac{\varOmega}{2}&Ja\\Ja^\dag&\omega a^\dag a-\frac{\varOmega}{2}
\end{array}\right).
\end{eqnarray}

Following a similar procedure for the DO, one can define the conserved number operator
\begin{align}
\mathcal{N}_{JC}&=\mathcal{V}N\mathcal{V}^\dag
=\begin{array}{cc}
 \left(
   \begin{array}{ccc}
     N & 0 \\
     0&N-1 \\
     \end{array}
 \right)
\end{array}
\end{align}
by the unitary
\begin{eqnarray}
\mathcal{V}=\left(\begin{array}{cc}1&0\\0&\frac{1}{\sqrt{N}}a^\dag
\end{array}\right).
\end{eqnarray}
The Hamiltonian can be diagonalized in two steps.
First, by the unitary, it is transformed into
\begin{eqnarray}
\mathcal{V}^\dag \mathcal{H}_{JC} \mathcal{V}
=\begin{array}{cc}
 \left(
   \begin{array}{ccc}
    \omega N+\frac{\varOmega}{2}& J\sqrt{N+1} \\
    J\sqrt{N+1}&\omega(N+1)-\frac{1}{2}\varOmega
     \end{array}
 \right).
\end{array}
\end{eqnarray}
Second, one can choose a  $2\times2$ operator in the  subspace with the same quantum number as
\begin{eqnarray}
&&e^{i\frac{\sigma_{2}}{2}\phi_N}  \mathcal{V}^\dag \mathcal{H}_{JC} \mathcal{V}  e^{-i\frac{\sigma_{2}}{2}\phi_N} \nonumber\\
&=\,&\omega(N+\frac{1}{2})+\sqrt{J^{2}( N+1)+\frac{(\varOmega-\omega)^{2}}{4}}\sigma_{3} ,  \ \ \
\end{eqnarray}
where $\phi_N$ is an operator function defined by $\tan\phi_N= 2 J\sqrt{N+1}/{(\varOmega -\omega)}$.
Hence, its eigenvalues are
  \begin{eqnarray}
\epsilon^{\pm}_n=\omega( n+\frac{1}{2})\pm\sqrt{J^{2}(n+1)+\frac{(\varOmega-\omega)^{2}}{4}},
\end{eqnarray}
 corresponding to the eigenstates
\begin{eqnarray}
|\psi_{n}^{\pm}\rangle = \mathcal{V} e^{-i\frac{\sigma_{2}}{2}\phi_N} |\pm\rangle\otimes | n\rangle,
\end{eqnarray}
where $n=0,1,2  ,\ldots $.
Two conserved quantities  and their shift operators
can be constructed as
\begin{eqnarray}
\mathcal{N}_{JC} &=&\mathcal{V}  e^{-i\frac{\sigma_{2}}{2}\phi_N}   N  e^{i\frac{\sigma_{2}}{2}\phi_N} \mathcal{V}^\dag , \nonumber \\
  {b}_{JC} &=&\mathcal{V}  e^{-i\frac{\sigma_{2}}{2}\phi_N}   a  e^{i\frac{\sigma_{2}}{2}\phi_N} \mathcal{V}^\dag ,\nonumber \nonumber\\
  {b}^\dag_{JC} &=&\mathcal{V}  e^{-i\frac{\sigma_{2}}{2}\phi_N}   a^\dag  e^{i\frac{\sigma_{2}}{2}\phi_N} \mathcal{V}^\dag ,\nonumber\\
\varSigma^{(3)}_{JC} &=&\mathcal{V}  e^{-i\frac{\sigma_{2}}{2}\phi_N}   \sigma_3  e^{i\frac{\sigma_{2}}{2}\phi_N} \mathcal{V}^\dag ,\nonumber \\
\varSigma^{(\pm)}_{JC} &=&\mathcal{V}  e^{-i\frac{\sigma_{2}}{2}\phi_N}   \sigma_{\pm} e^{i\frac{\sigma_{2}}{2}\phi_N} \mathcal{V}^\dag .
\end{eqnarray}
 By following the same steps to study the DO, one can show an $SO(4)$ algebraic structure in  the JC model.

In summary, by extending the approach to study the 2D Dirac system with a spin symmetry,
we present a unitary which  transforms the number operator to a conserved quantity of the 1D DO.
Based on this result, one can diagonalize the  DO Hamiltonian, and obtain the two conserved quantities together with  their shift operators.
These operators show an  $SO(4)$ algebra connecting the eigenstates of the Dirac oscillator.
By a similar procedure,   we also show the $SO(4)$ algebra in the JC model.

Further research on this topic in the two following directions would be interesting.
First,  whether one can extend the conserved number operator to the two- or three-dimensional Dirac oscillators, and derive a deformed isotropic harmonic oscillator,  is a natural question.
Second, it is fascinating to consider  the algebraic of the multiphoton JC model \cite{huai2000multiphoton}.
More specifically, we look forward to obtaining a polynomial generalization of the $SO(4)$ algebra \cite{PhysRevA.80.054102} in it.

\emph{Acknowledgments. }This work was supported by the National Natural Science Foundation of China (Grant Nos. 11675119, 11575125, and  11105097).

\end{document}